\documentclass{article}
\usepackage{spconf,amsmath,graphicx}
\usepackage{amsfonts} 
\usepackage{IEEEtrantools} 
\usepackage{balance} %
 
\usepackage{epstopdf}  
\usepackage{verbatim}  
\usepackage{color} 
\usepackage{pifont} 
\usepackage{pmat} 
\usepackage{booktabs} 
\usepackage{multirow}
\usepackage{bigstrut}
\usepackage{array}    
\usepackage{esvect}  
\usepackage{pgfplots}  
\usetikzlibrary{plotmarks}
 \usepackage{url}

\DeclareMathOperator{\Diag}{\mathsf{bdiag}}
\DeclareMathOperator{\Stack}{\mathsf{stack}}

\DeclareMathOperator{\An}{\mathbf{A}}
\DeclareMathOperator{\Vn}{\mathbf{V}}

\DeclareMathOperator{\Hn}{\mathbf{H}}

\DeclareMathOperator{\yn}{\mathbf{y}}
\DeclareMathOperator{\Gn}{\mathbf{G}}
\DeclareMathOperator{\Un}{\mathbf{U}}

\DeclareMathOperator{\Tn}{\mathbf{T}}
\DeclareMathOperator{\nn}{\mathbf{n}}
\DeclareMathOperator{\0}{\mathbf{0}}

\DeclareMathOperator{\enters}{\mathbb{Z}^+}

\newcommand{\espai}{$\text{}$ \\} 
\newcommand{\hespai}{$\!\!\!\!\!\!$} 
\newcommand{\Cmat}[2]{\!\in\!\mathbb{C}^{#1 \times #2}}

\newcommand{\Span}[1]{\mathsf{span} \!\left(  #1  \right)}
\newcommand{\RSpan}[1]{\mathsf{rspan} \left( #1 \right)}
\newcommand{\Rank}[1]{\mathsf{rank} \big( #1 \big)}
\newcommand{\Dim}[1]{\mathsf{dim} \left( #1 \right)}

\newcommand{\textboxx}[3]{\node[font=\small ] at (axis cs: #1,#2) {#3};} 

\renewcommand{\IEEEQED}{\IEEEQEDopen} 
\renewcommand{\IEEEproofindentspace}{0em}

\newtheorem{theorem}{Theorem}

\title{Retrospective Interference Alignment \\ for the 3-user MIMO Interference Channel with delayed CSIT}
\name{Marc Torrellas, Adrian Agustin, Josep Vidal\thanks{This work has been done in the framework of the projects TEC2010-19171/TCM, CONSOLIDER INGENIO CSD2008-00010 COMONSENS, 2009SGR1236
(AGAUR) of the Catalan Administration, and TROPIC FP7 ICT-2011-8-318784 project, funded by the European Commission. Draft version of the accepted manuscript at IEEE ICASSP 14.
}}
\address{Universitat Polit\`ecnica de Catalunya (UPC), Barcelona \\
\{marc.torrellas.socastro, adrian.agustin, josep.vidal\}@upc.edu}
\begin{document}
%
\copyright 2014 IEEE. Personal use of this material is permitted. Permission from IEEE must be 
obtained for all other uses, in any current or future media, including 
reprinting/republishing this material for advertising or promotional purposes, creating new 
collective works, for resale or redistribution to servers or lists, or reuse of any copyrighted 
component of this work in other works.

\balance
\newpage

\maketitle
\begin{abstract}
The degrees of freedom (DoF) of the 3-user multiple input multiple output interference channel (3-user MIMO IC) are investigated where there is delayed channel state information at the transmitters (dCSIT). We generalize the ideas of Maleki et al. about {\it Retrospective Interference Alignment (RIA)} to be applied to the MIMO IC, where transmitters and receivers are equipped with $(M,N)$ antennas, respectively. 
We propose a two-phase transmission scheme where the number of slots per
phase and number of transmitted symbols are optimized by solving a maximization problem.
Finally, we review the existing achievable DoF results in the literature as a function of the ratio between transmitting and receiving antennas $\rho=M/N$. The proposed scheme improves all other strategies when 
$\rho \in \left(\frac{1}{2}, \frac{31}{32} \right]$.
\end{abstract}
\section{Introduction}
\label{sec:intro}

In recent years, IA has become one of the most promising tools to analyze interference networks at high signal to noise ratio (SNR)\cite{Maddah-Ali2008,CJ}. This technique shows that each user of $K$ interfering pairs communicating simultaneously can get ``half of the cake'', i.e each user achieves half the DoF as compared to the single-user case. The concept relies on the fact that each user does not care about other user's messages. Therefore, the transmitted signals are designed in such a way that they are enclosed on a common subspace at the non-intended receivers, which is disjoint from the subspace generated by the intended signals. Nevertheless, IA-based schemes require instantaneous CSIT, an assumption not always valid in wireless cellular networks.
  
In this regard, a new framework for the analysis of the MISO Broadcast Channel when the CSIT is perfect but delayed (referred to as dCSIT), was introduced in \cite{MAT}. This is an intermediate situation among perfect CSIT \cite{SubspaceAlignmentChains} and no CSIT \cite{VazeNoCSI_Tr}. Recent results \cite{MalekiRIA,Vaze2IC,Ghasemi2011,Abdoli_IC,KUMIC} have analyzed the DoF of broadcast and interference scenarios in dCSIT conditions. In all these approaches, the IA concept is generalized for a transmission carried out in multiple phases, where the signals are aligned along the space-time domain. 

In this work we investigate the achievable DoF of the 3-user MIMO IC
presented in Fig.1 when transmitters and receivers are equipped with $M$ and $N$ antennas, respectively. We generalize the 2-phase scheme introduced in \cite{MalekiRIA} for the 3-user SISO IC to the MIMO setting, performing the RIA in an alternative way. The number of slots per phase and the number of transmitted symbols are optimized to maximize the achievable DoF. We get a closed-form solution for such parameters and derive the achieved DoF as a function of the ratio $\rho=M/N$. Moreover, using the obtained results and those existing in the literature the best-known achievable DoF are found as a function of $\rho$. The proposed scheme stands as the best one when $\rho \in \left(\frac{1}{2}, \frac{31}{32} \right]$.


{\bf Notation}: Boldface and lower case fonts denote column vectors ($\mathbf{x}$). Boldface and upper case is used for matrices ($\mathbf{X}$). 
$(\cdot)^H$, and $\0$ are the transpose and conjugate operator, and the all-zero matrix, respectively. Some predefined vector and matrix operations are detailed below:
\begin{IEEEeqnarray}{c}
\begin{matrix}
\Stack
\left( \An,\mathbf{B}\right) =
\begin{pmatrix}
\An \\
\mathbf{B} \\
\end{pmatrix} 
&
\Diag
\left( \An,\mathbf{B} \right) =
\begin{pmatrix}
	\mathbf{A} & \0 \\
	\0 & \mathbf{B} \\
\end{pmatrix}
\end{matrix} \nonumber
\end{IEEEeqnarray}
Further, $\Span{\An}$ defines the subspace generated by all linear combinations of the columns of $\An$, ${\RSpan{\An}=
\mathsf{span} \big( \An^T \big) }$, and
${\Rank{\An}=\Dim{\Span{\An}} =\Dim{\RSpan{\An}} }$. 
Further, $\mathbb{C}$ and $\enters$ denote the set of complex numbers, and positive integers, respectively.
Finally, all indexes in this work are assumed to be in the set $\{1,2,3\}$, thus applying the modulo-$3$ operation when necessary.

\begin{figure}[h]
\begin{minipage}[]{1\linewidth}
  \centering 
  \centerline{\includegraphics[width=0.5\linewidth]{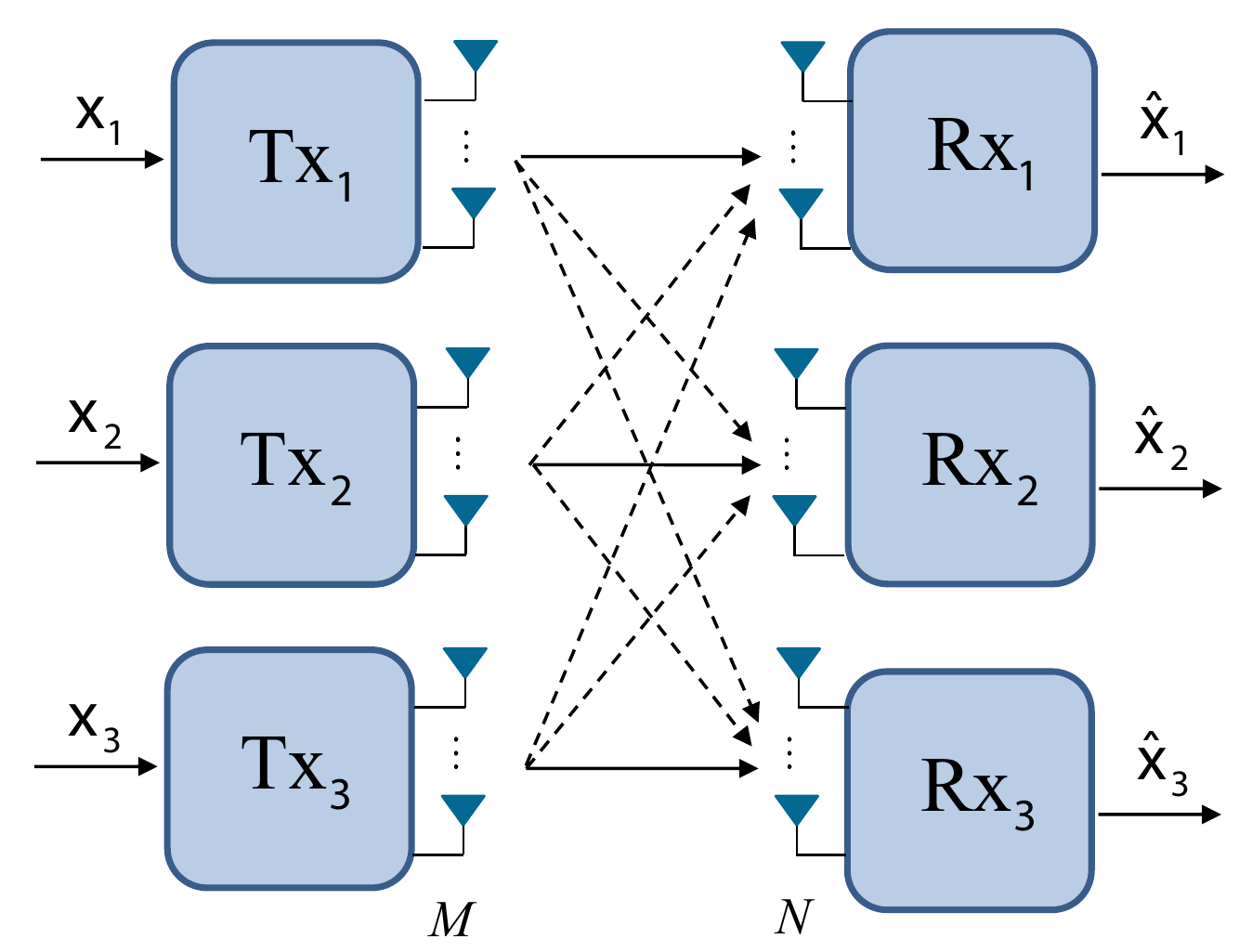}}
\end{minipage}  
%
\caption{The $3$-user MIMO IC, with $(M,N)$ antennas at the transmitters and receivers, respectively.}
\label{fig:scenarios}
\end{figure}

\section{SYSTEM MODEL}
\label{sec:SystemModel}

The 3-user MIMO IC scenario consists of 3 transmitter-receiver pairs interfering with each other, see Fig.\ref{fig:scenarios}. Each transmitter is equipped with $M$ antennas, and delivers $b$ independent symbols to its associated receiver, equipped with $N$ antennas.
The transmission is carried out in $2$ phases of duration $W_1$ and $W_2$ slots, with $W=W_1+W_2$ denoting the total number of slots for the communication. All transmitters are active during all slots. The first phase, denoted by {\it Interfering Sensing phase (IS phase)}, is employed by users for sensing the interference. This knowledge is then communicated to transmitters. During the second phase, denoted by {\it RIA phase}, transmitters use such information to design the transmit precoders performing RIA. Further details are specified in Section \ref{sec:Maleki}.

The output at the $j$th receiver during each time slot is described by:
\vspace{-3mm}
\begin{IEEEeqnarray}{c}
\vspace{-1mm}
\yn_j^{\left(p,s\right)} = \sum_{i=1}^{3} \Hn_{j,i}^{\left(p,s\right)}
 \Vn_i^{\left(p,s\right)} \mathbf{x}_i + \nn_j^{(p,s)}
\label{eq:SystemModel}
\end{IEEEeqnarray}
where $\Vn_i^{\left(p,s\right)}\Cmat{M}{b}$ is the precoding matrix used by the $i$th user at time slot $(p,s)$, i.e the $s$th time slot of the $p$th phase, $\mathbf{x}_i \Cmat{b}{1}$ contains the intended symbols to the $i$th receiver, $\nn_j^{(p,s)} \Cmat{N}{1}$ is the zero-mean unit-variance Additive White Gaussian Noise (AWGN) term, and $\Hn_{j,i}^{\left(p,s\right)} \Cmat{N}{M}$ is the channel matrix containing the gains from antennas of the $i$th transmitter to the $j$th receiver. The channels are assumed to be flat block-fading and i.i.d. across time, i.e. the channel coefficients are completely uncorrelated among slots.

Since the transmission is carried out over multiple time
slots, the channel model in (\ref{eq:SystemModel}) can be alternatively written by grouping the magnitudes of the different phases:  
\begin{IEEEeqnarray}{c}
\label{eq:SystemModelextended}
\begin{matrix}
\mathbf{r}_j =   \mathbf{Z}_j \left(
\sum_{i=1}^{3} \mathbf{H}_{j,i} \mathbf{V}_i \mathbf{x}_i + \nn_j  \right) \\[2mm]
\mathbf{H}_{j,i}^{(p)} = \Diag
\left(
\begin{matrix}
	\Hn_{j,i}^{\left(p,1\right)} &
	\Hn_{j,i}^{\left(p,2\right)} &\ldots &
	\Hn_{j,i}^{\left(p,W_p\right)}
\end{matrix} 
\right)
\\[2mm]
 \mathbf{V}_i^{(p)} = \Stack
\left(
\begin{matrix}
	{\Vn_i^{\left(p,1\right)}}
	&{\Vn_i^{\left(p,2\right)}} 
	& \ldots & {\Vn_i^{\left(p,W_p\right)}}
\end{matrix} 
\right) \\[2mm]
\mathbf{H}_{j,i} = \Diag \left( \mathbf{H}_{j,i}^{(1)} , \mathbf{H}_{j,i}^{(2)}\right) \, 
, \Vn_{i} = \Stack \left( \Vn_{i}^{(1)} , \Vn_{i}^{(2)} \right)
\end{matrix}
\end{IEEEeqnarray}
\\
for $p=1,2$, where $\mathbf{V}_i\Cmat{MW}{b}$, $\mathbf{H}_{j,i} \Cmat{NW}{MW}$, $\mathbf{V}_i^{(p)}\Cmat{MW_p}{b}$ and ${\mathbf{H}_{j,i}^{(p)} \Cmat{NW_p}{MW_p}}$. Finally, ${\mathbf{r}_j\Cmat{b}{b}}$ represents the signal processed with ${\mathbf{Z}_j\Cmat{b}{WN}}$, the zero-forcing receiving filter removing the received interference. The transmit and receive filter design should satisfy the following two conditions:
\begin{IEEEeqnarray}{c}
\Rank{ \mathbf{Z}_{j} \mathbf{{H}}_{j,j} \mathbf{{V}}_j } = {b} \label{eq:noInterferencea}
\\
\mathbf{Z}_{j} \mathbf{{H}}_{j,i} \mathbf{{V}}_i  = \0 , \quad i \neq j
\label{eq:noInterferenceb}
\end{IEEEeqnarray}
where (\ref{eq:noInterferenceb}) states that the receiving filter projects the received signal onto the orthogonal-to-interference subspace, while (\ref{eq:noInterferencea}) ensures that the obtained subspace contains $b$ independent linear combinations of the desired symbols. Notice that these two conditions are ensured if the desired and interference signals are linearly independent. In such a case, the achievable DoF per user are given by $\hat{d}=b / W$.

In the following we describe how the CSI is computed at the receiver side at the end of the IS phase and reported to transmitters to be used during the RIA phase. To this end, let define $ {\Tn_{j,i} = \Un_{j,i} \Hn_{j,i}^{\left(1\right)}  \! \Vn_{\!i}^{\left(1\right)} \Cmat{(NW_1 - b)}{b}}$ as the matrix containing the interference received at the $j$th receiver from the $i$th transmitter. Note that matrix ${\Un_{j,i} \Cmat{(NW_1-b)}{NW_1}}$, ${\forall i \neq j }$ is a linear filter such that
\scalebox{0.95}{
\begin{minipage}{\linewidth}
\begin{IEEEeqnarray*}{c}
\begin{bmatrix}
 \Un_{j,j+1} \\ \Un_{j,j-1}
\end{bmatrix}
 \left[
      \Hn_{j,j+1}^{\left(1\right)} \! \! \Vn_{\!\!j+1}^{\left(1\right)}  , 
      \Hn_{j,j-1}^{\left(1\right)} \! \! \Vn_{\!\!j-1}^{\left(1\right)} 
 \right] 
\! = \! \!
\begin{bmatrix}
   \Tn_{j,j+1} & \0 \\ 
  \0 & \Tn_{j,j-1}
 \end{bmatrix}
\end{IEEEeqnarray*}
\end{minipage}
}
\vspace{-10mm}
\begin{align}
\label{eq:partialCancellation} 
\end{align}
\vspace{0.5mm}

Therefore, we assume that during the RIA phase the $i$th transmitter knows $\left\{ S_{i+1,i} , S_{i-1,i} \right\}$, with $ S_{j,i} = \RSpan{\Tn_{j,i}}$, i.e. each transmitter has access to the subspaces spanned by the interference caused by itself at each unintended receiver during the IS phase.

\section{Main Result}
\label{sec:mainResults}

\begin{theorem}
 For the 3-user MIMO IC with dCSIT, where the transmitters and receivers are equipped with (M,N) antennas, respectively, the following DoF per user can be achieved for each value of $\rho = \frac{M}{N}$:
 \label{th:innerbound} 
\end{theorem}
\vspace{-3mm}
\begin{table}[h]
\begin{center}
\scalebox{1}{%
%
%
%

\begin{tabular}{ c | c | c | c | c | c | c}
    Case & $\rho$ & $\hat{d}$ & \hespai & Case & $\rho$ & $\hat{d}$ \\


\hline  \bigstrut
    \multirow{2}{*}{\hspace{2.5mm}\vspace{-3.5mm}A} & $\left[ 0 , \frac{1}{3} \right) $ & $M$  & \hespai & C & $\left[\frac{31}{32} , \frac{18}{13} \right) $ & $\frac{12}{31}\min(M,N) $ \\[0.1pt]
    \cline{5-7}  \bigstrut
     & $\left[\frac{1}{3} ,\frac{1}{2}\right]$ & $\frac{N}{3}$ & \hespai & \multirow{2}{*}{\vspace{-3.5mm}D} & $ \left[ \frac{18}{13} , 2 \right) $ & $\frac{2}{3} \frac{MN}{M+N}$ \\
     \cline{1-3} \bigstrut
    \multirow{2}{*}{\hspace{2.5mm}\vspace{-3.5mm}B} & $\left( \frac{1}{2} , \frac{3}{5}\right]$ & $\frac{MN}{M+N}$ & \hespai & & $\left[ 2 , 3 \right)$ & $\frac{4}{9}N $ \\
    \cline{5-7} \bigstrut
    & $\left(\frac{3}{5} , \frac{31}{32}\right] $ & $\frac{3}{8}N$ & \hespai & \vspace{1mm}E & $\left[ 3, \infty \right) $ & $\frac{1}{2}N$
\end{tabular} 

}
\end{center}
\vspace{-4mm}
\end{table} 

\begin{IEEEproof}  
 The case A is tackled by using the CSIR only optimal scheme, i.e. by means of zero-forcing concepts at the receivers. Since these inner bounds are tight when there is full CSIT \cite{SubspaceAlignmentChains}, they are also tight for the case of dCSIT. The DoF for region B are derived later in the present work. In case C, a scaled version of the scheme presented in \cite{Abdoli_IC} for the SISO setting is applied, by turning off the additional antennas at each transmitter (if $M>N$) or receiver (if $M<N$). Moreover, for the case D we apply the 2-user IC optimal scheme in \cite{Vaze2IC} by scheduling two users for each considered transmission period\footnote{Since we are interested in the symmetric DoF, the DoF achieved in the 3-user MIMO IC can be derived by using TDMA of the 3 possible 2-user MIMO ICs. Hence, the DoF per user are those achieved in each 2-user MIMO IC multiplied by $\frac{2}{3}$.}. Finally, the inner bound for $\rho \geq 3$ is achieved by using the 2-phases scheme (or a scaled version of it) for the $K$-user MIMO IC proposed in \cite{KUMIC}.
\end{IEEEproof}   

\section{Transmission scheme for case B}
\label{sec:Maleki}


We extend the strategy of \cite{MalekiRIA} to the MIMO case, and with general values for $b, W_1$ and $W_2$. For a richer understanding, our approach is presented in an alternative fashion, inspired by the methodology in \cite{Abdoli_IC}. As a result, we build a maximization problem with linear constraints that allows to design such parameters. Finally, an analytical solution is obtained, proving Theorem \ref{th:innerbound}-Case B.

\subsection{Interfering Sensing phase}
\label{sec:trainPhase}
Since there is no CSIT of the current channels, during $W_1$ time slots each transmitter employs predefined precoders that are known by all the users, thus each receiver obtains

\begin{IEEEeqnarray}{c}
\!\!\!\!\yn_j^{\left(1\right)} \!\!= \! \Hn_{j,j}^{\left(1\right)} \!
 \Vn_j^{\left(1\right)} \!\! \mathbf{x}_j \! + \!  
 \left[
      \Hn_{j,j+1}^{\left(1\right)} \! \! \Vn_{\!\!j+1}^{\left(1\right)}  , 
      \Hn_{j,j-1}^{\left(1\right)} \! \! \Vn_{\!\!j-1}^{\left(1\right)} 
 \right] \!\!
 \begin{bmatrix}
  \mathbf{x}_{j+1} \\ \mathbf{x}_{j-1}
 \end{bmatrix}
 \! \!  + \! \nn_j^{(1)}
\end{IEEEeqnarray}

\subsection{Retrospective Interference Alignment phase}
\label{sec:alignPhase}

By using the IS phase report, each transmitter aims to align the signals transmitted during the RIA phase, with the previous received interferences at {\it both} interfered receivers. Hence, the $i$th transmitted signal should satisfy the following conditions:

\begin{IEEEeqnarray}{c}
\RSpan{ 
      \Hn_{i+1,i}^{(2,s)}\! \mathbf{V}_{i}^{(2,s)} } 
      \subseteq 
      S_{i+1,i}  \\
      \RSpan{ 
      \Hn_{i-1,i}^{(2,s)}\! \mathbf{V}_{i}^{(2,s)} } 
      \subseteq 
      S_{i-1,i} 
\end{IEEEeqnarray}

An easy way to ensure this is to set
\begin{IEEEeqnarray}{c}
 \mathbf{V}_{i}^{(2,s)}  =  \sigma_i
 \begin{bmatrix}
      \mathbf{T}_{i}     \\
      \0
 \end{bmatrix}
\label{eq:RIA_simple} \\[1mm]
\RSpan{\Tn_i} = S_i = S_{i+1,i}  \cap S_{i-1,i} 
\label{eq:intersection}
\end{IEEEeqnarray}
where $\sigma_i$ is adjustable to ensure the transmission power constraint, and $\mathbf{T}_i \Cmat{\Dim{S_i}}{b}$.
This precoding matrix is used during $W_2$ slots, ensuring that each transmitter does not increase the interference caused at each non-intended receiver, while delivering fresh linear combinations of desired symbols to each intended receiver.

%

\subsection{Feasibility}
\label{sec:feasibility}

The linear constraints that restrict $b$, $W_1$, and $W_2$ for the proposed precoding scheme are derived in the following.
\\[1.5mm]
1) {\it Existence of $\Un_{j,i}$}: Each receiver will be able to compute a solution for $\Un_{j,i}$ in (\ref{eq:partialCancellation}) whenever $b \leq 
\min\left(MW_1, NW_1-1\right)$.
\\[1mm]
2) {\it Intersection subspace}: We should guarantee the existence of each subspace $S_i$ in (\ref{eq:intersection}), whose dimension is given by
\footnote{For any subspaces $A$ and $B$, $\Dim{{A}\cap B}= \Dim{A}+\Dim{B}-\Dim{A+B}$ holds. \cite{Grassmann}}:
\begin{IEEEeqnarray}{c}
 \Dim{S_i}= 2 \min\left( NW_1 - b,b \right) - b > 0
 \label{eq:dimIntersection}
\end{IEEEeqnarray} 
3) {\it Receiver space-time dimensions}: Each receiver should have enough space-time dimensions to allocate all the desired and interference signals without space overlapping. First, notice that the interference received during the IS phase occupies at most $NW_1$ dimensions. This subspace remains the same after the RIA phase, since all the interference is aligned. On the other hand, the desired signals occupy at most $b$ dimensions at each receiver. Hence, we must have
\begin{IEEEeqnarray}{c}
\underbrace{b_{ }}_{\text{desired dim.}} + \underbrace{NW_1}_{\text{interference dim.}} \leq \underbrace{NW_1 + N W_2}_{\text{total dimensions}} \Rightarrow
b \leq N W_2
\label{eq:constraintEspai}
\end{IEEEeqnarray} 
4) {\it Rank of desired signals after zero-forcing}: each receiver needs $b$ linearly independent combinations of the desired signals after the interference is zero-forced. This can be guaranteed whenever the desired and interference signals are linearly independent, or equivalently, if the matrix
\begin{IEEEeqnarray}{c}
\Gn_j \!= \!
\begin{bmatrix}
 \Gn_j^{(1)} \\[1mm]
 \Gn_j^{(2)}
\end{bmatrix}
\!=\! 
\begin{bmatrix}
  \Hn_{j,j}^{(1)} \Vn_j^{(1)} & \Hn_{j,j+1}^{(1)} \Vn_{j+1}^{(1)} & \Hn_{j,j-1}^{(1)} \Vn_{j-1}^{(1)}
  \\[1mm]
  \Hn_{j,j}^{(2)} \Vn_j^{(2)} & \Hn_{j,j+1}^{(2)} \Vn_{j+1}^{(2)} & \Hn_{j,j-1}^{(2)} \Vn_{j-1}^{(2)}
\end{bmatrix}
\end{IEEEeqnarray}
is full rank. Due to space limitation, we give some intuition about the proof, and a sufficient condition to ensure that $\Gn_j$ is full rank.
Consider the first $NW_1$ rows of $\Gn_j$, i.e. the rows corresponding to the IS phase, denoted by $\Gn_j^{(1)}$. Since the precoding matrices are independent and randomly chosen, $\Gn_j^{(1)}$ becomes full rank with probability equal to 1. On the other hand, let denote by $\Gn_j^{(2)}$ the last $NW_2$ rows of $\Gn_j$, corresponding to the signals received during the RIA phase. Since $\Hn_{j,j}^{(2)} \Vn_j^{(2)}$ is independent of the channels in $\Gn_j^{(1)}$ (only depends on channels terminating at other receivers), the rows in $\Gn_j^{(1)}$ and $\Gn_j^{(2)}$ become linearly independent. Thus, we need only to show that $\Gn_j^{(2)}$ is full rank. To this end, remember that $\Rank{\Vn_i^{(2)}}=\Dim{S_i}$, see (\ref{eq:RIA_simple}) and (\ref{eq:dimIntersection}). Also, note that $\Rank{\Hn_{j,i}^{(2)} \Vn_i^{(2)}}=\Rank{\Vn_i^{(2)}}$, since the product of matrices cannot increase the rank, and due to $\Vn_i^{(2)}$ is designed independently of the current channel state. As a result, we can state that the rank of $\Gn_j^{(2)}$ can be computed as
\begin{IEEEeqnarray}{c}
 \begin{matrix}
  \Rank{\Gn_j^{(2)}}=\sum_{i=1}^{3} \Rank{\Hn_{j,i}^{(2)} \Vn_i^{(2)}} \geq b \\[3mm]
3 \min\left( 2NW_1 - 3b,b \right) \geq b \Rightarrow
5b \leq 3 N W_1
 \end{matrix}
\label{eq:rankConstraint}
\end{IEEEeqnarray}

\subsection{DoF optimization problem}
\label{sec:MalekiMIMO}
The constraints derived in the previous section are collected to formulate the following DoF maximization problem:
\begin{IEEEeqnarray}{c l}
 \underset{ \left\{b,W_1,W_2 \right\} \in \enters}{\text{maximize}} \quad  & \frac{b}{W_1+W_2} \label{eq:objfunc} \\[2mm]
s.t. & b \leq \min\left(MW_1,NW_1-1\right) \label{eq:maxProblema} \\
     & 2 \min\left( NW_1 - b,b \right) - b > 0 \label{eq:maxProblembb} \\
     & 5b \leq 3 N W_1 \label{eq:maxProbleme} \\
     & b \leq N W_2 \label{eq:maxProblemd}
\end{IEEEeqnarray}
For any given value of $b$, the objective function in (\ref{eq:objfunc}) is strictly decreasing with $W_1$ and $W_2$, i.e. their optimum values are their minimum feasible values. Therefore, since $W_2$ only appears in (\ref{eq:maxProblemd}), the optimum value $W_2^*$ is given by
\begin{IEEEeqnarray}{c}
{W_2^* =  \Big\lceil \dfrac{b}{N}} \Big\rceil
\end{IEEEeqnarray}
On the other hand, the optimum value $W_1^*$ will be equal to the minimum feasible value for $W_1$ satisfying (\ref{eq:maxProblema})-(\ref{eq:maxProbleme}). This is formulated as follows: 
\begin{IEEEeqnarray}{c}
W_1^* \! = \max \!\left( {\frac{b}{M}}, \frac{b+1}{N}, {\frac{3b+1}{2N}}, {\frac{5}{3}\frac{b}{N}} \right)\!, \text{iff } \, {W_1^* \!\leq \!\frac{2b-1}{N}} 
%
\end{IEEEeqnarray}
One optimal solution for $b^*,W_1^*$ and $W_2^*$ for the two regions configuring case B (see Theorem 1) is shown below:

\begin{table}[h]
\begin{center}
\begin{tabular}{c | c | c | c | c}
  &  $b^*$  & $W_1^*$ & $W_2^*$ & $\hat{d}$  \\
\cline{1-5} \bigstrut
$\,\frac{1}{2} < \rho \leq \frac{3}{5} $ & $MN $ & $N$ & $M$ & $\frac{MN}{M+N}$ \\
\cline{1-5} \bigstrut
 $\frac{3}{5} \leq \rho \leq \frac{31}{32} $ & $3N$ & 5 & 3 & $\frac{3}{8} N$ \\
\end{tabular} 
\end{center}
\vspace{-5mm}
\label{tab:summary_M=3}
\end{table} 
\vspace{-3mm}
\espai
where $\rho= \frac{M}{N}$, and we have applied that $b,W_1,W_2 \in \enters$.

%
%
%

          

\section{Results}
\label{sec:results}

\begin{figure}[h]
\begin{minipage}[b]{1\linewidth}
  \centering   

\tikzstyle{every pin}=[
font=\footnotesize,pin distance=0.6cm,inner sep=1pt]

\begin{tikzpicture}
\begin{axis}[
  ymin=0,ymax=0.6,xmin=0,xmax=3.5,
  xmajorgrids,
   ymajorgrids,
   grid style={dashed, gray!30},
  ylabel style={at={(0.07,0.5)},rotate=-90},
  width=0.7\linewidth, 
   xtick={0.333,0.5,0.6,0.968,1.384,2,3},
   xticklabels={$\frac{1}{3}$,$\frac{1}{2}$,$\frac{3}{5}$,$\frac{31}{32}$,$\frac{18}{13}$,$2$,$3$},
   ytick={0.333,0.3874,0.444,0.5,0.5455},
   yticklabels={$\frac{1}{3}$,$\frac{12}{31}$,$\frac{4}{9}$,$\frac{1}{2}$,$\frac{6}{11}$},
  font=\scriptsize,
  xlabel=$\rho$,
  ylabel=$\frac{\hat{d}}{N}$,
  legend style={at={(axis cs: 3.25,0.05)},anchor=south east,font=\scriptsize}],
  ]

  \addplot[red, densely dashed,  thick,domain=0:1/3]{x};
  \addlegendentry{Outer bound};
  \addplot[forget plot, red, densely dashed,  thick,domain=1/3:1/2]{1/3};
  
  \addplot[forget plot, red,densely dashed ,  thick,domain=1/2:3/5]{2/3*x};
  \addplot[forget plot, red, ,  thick,domain=3/5:2/3]{2/5};
  
  \addplot[forget plot, red, densely dashed,  thick,domain=2/3:5/7]{3/5*x};
  \addplot[forget plot, red, densely dashed,  thick,domain=5/7:3/4]{3/7};
  
  \addplot[forget plot, red, densely dashed,  thick,domain=3/4:7/9]{4/7*x};
  \addplot[forget plot, red, densely dashed,  thick,domain=7/9:4/5]{4/9};

  
  \addplot[forget plot, red, dotted,  thick,domain=4/5:6/5, samples=3, samples y = 3]{0.253*x + 8/33};
  \addplot[forget plot, red, densely dashed,  thick,domain=6/5:3.5]{6/11};
  
  \addplot[forget plot,black, thick,domain=3:3.5]{1/2};
  \addplot[forget plot,black, thick,domain=2:3]{4/9};
  \addplot[forget plot,black, thick,domain=18/13:2]{2/3*x/(x+1)};
  
  \addplot[forget plot,black, thick,domain=1:18/13,samples=50]{12/31};
  
  \addplot[forget plot,black, thick,domain=31/32:1,samples=50]{12/31*x};
  
  \addplot[forget plot, black, thick,domain=31/36:1,samples=50]{12/31*x};
    
  \addplot[forget plot,blue,densely dotted,very thick,domain=1/2:3/5,samples=50]{x/(x+1)};
  
  \addplot[blue,densely dotted,very thick,domain=3/5:31/32,samples=50]{3/8};
  \addlegendentry{New inner bounds};
  
  \addplot[black, thick,domain=1/3:1/2]{1/3};
  \addlegendentry{Previous inner bounds};
  
  \addplot[forget plot,black, thick,domain=1/3:31/36]{1/3};
  
  \addplot[forget plot,black, thick,domain=0:1/3]{x};

  \node[coordinate,pin={[pin distance=0.32cm]below:{$\frac{2}{3}\frac{\rho}{\rho+1}$}}] at (axis cs: 1.7,0.418) {};
  \node[coordinate,pin={[pin distance=0.35cm]-55:{$\frac{12}{31} \rho $}}] at (axis cs: 0.93,0.36) {};
  \node[coordinate,pin={[pin distance=0.5cm,inner sep = 0.2pt]25:{$\frac{3}{8}$}}] at (axis cs: 0.77,0.375) {};
  \node[coordinate,pin={[pin distance=0.45cm,inner sep = 0pt]-70:{$\frac{\rho}{\rho+1}$}}] at (axis cs: 0.55,0.35) {};
  \textboxx{0.26}{0.18}{$\rho$}


  \draw [fill=white] (axis cs: 0,0.56) rectangle (axis cs: 0.5,0.6);    
  \textboxx{0.25}{0.577}{A} 
  
  \draw [fill=white] (axis cs: 0.5,0.56) rectangle (axis cs: 0.967,0.6);    
  \textboxx{0.75}{0.577}{B}
  
  \draw [fill=white] (axis cs: 0.968,0.56) rectangle (axis cs: 1.3846,0.6);    
  \textboxx{1.192}{0.577}{C}
  
  \draw [fill=white] (axis cs: 1.3846,0.56) rectangle (axis cs: 3,0.6);    
  \textboxx{2.192}{0.577}{D}
  
  \draw [fill=white] (axis cs: 3,0.56) rectangle (axis cs: 3.5,0.6);    
  \textboxx{3.25}{0.577}{E}
  
\end{axis}
\end{tikzpicture}
%
\end{minipage}
\vspace{-7mm}       
\caption{Different regions for the 3-user MIMO IC with dCSIT defined in Theorem 1 for each antenna setting.}
\label{fig:innerbound} 
\end{figure}
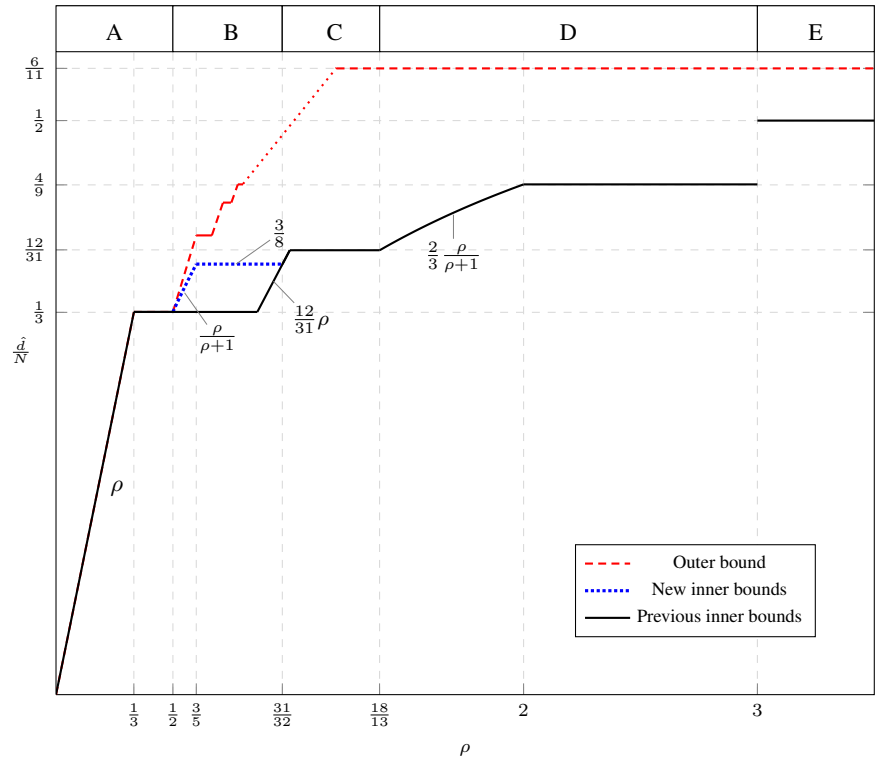

In Fig. \ref{fig:innerbound} we compare different strategies known in the literature
together with the proposed approach in terms of $\frac{\hat{d}}{N}$ as a
function of $\rho$ (see Theorem 1). We also present in dashed lines a simple
outer bound obtained as the minimum of 3 outer bounds: {\it i)}
assuming transmitter/receiver cooperation, the 3-user MIMO
IC can be reduced into a 2-user MIMO IC or a 3-user MIMO
BC with the corresponding antennas, and therefore the outer bounds in \cite{MAT,Vaze2IC} for such channels apply, {\it ii)} outer
bound assuming full CSIT for this channel \cite{SubspaceAlignmentChains}. We can observe
5 different regions for the inner bound. Despite the proposed inner bounds are not shown to be tight (this occurs only for $\rho \leq \frac{1}{2}$), we draw some conclusions for each case:
\begin{itemize}
 \item A: The receivers allocate the received signals in 3 disjoint subspaces of dimension $\min(\frac{N}{3},M)$ each, and remove completely all the interference by applying a ZF receiver, given the high number of receive antennas. 
 \item B: when there are enough antennas at each receiver, a ZF receiver uncouples the effect of the interference received from each transmitter. Based on that, the second phase follows RIA concepts by exploiting the residual interferences to align the transmitted signals.
  \item C: the $M-N$ additional antennas at the transmitters seem to be redundant, and a scaled version of the scheme in \cite{Abdoli_IC} derived for the SISO case is applied. The idea of antenna redundancy was already reported in \cite{SubspaceAlignmentChains} for this channel with perfect CSIT. Up to the author's knowledge, the technique presented in \cite{Abdoli_IC} has not been generalized to the MIMO case.
 \item D: This result is obtained by scheduling only two users
for each transmission period, who develop a 2-phases transmission based on RIA concepts. It seems that a 2-phases scheme with 3 active users (as proposed for case E) could outperform the present inner bound for this region.
 \item E: the best strategy is to apply the 2-phase scheme presented in \cite{KUMIC}. This is similar to what occurs for the 2-user MIMO IC (see \cite{Vaze2IC}) when the number of antennas at the transmitters is high enough.

\end{itemize}

\vspace{-1mm}
\section{Conclusions}
\label{sec:conclusions}

This work has investigated the DoF of a 3-user MIMO IC when the transmitters have dCSIT only. The main contribution is the generalization of the ideas in \cite{MalekiRIA} to the MIMO case. The achieved DoF values are greater than the best previously known achievable DoFs when 
$\frac{M}{N} \in \left(\frac{1}{2}, \frac{31}{32} \right]$. Future work may be oriented to optimize the resources (i.e. number of transmitted symbols per user, and number of slots per phase) for any antenna configuration.
\newpage
\balance
\bibliographystyle{IEEEbib}
\bibliography{3UMIC}

\end{document}